\UseRawInputEncoding
\documentclass[prd,twocolumn,showpacs,floatfix,amsmath,amssymb,floatfix]{revtex4}
\usepackage{graphicx,color,dcolumn,booktabs,bm}
\usepackage{longtable,lscape}
\usepackage{txfonts}
\usepackage{overpic}
\usepackage{amssymb}
\usepackage{indentfirst}
\usepackage{feynmf}   
\usepackage{slashed}  
\usepackage{cases}
\usepackage{color}
\usepackage{multirow}
\usepackage{epstopdf}
\usepackage{graphicx,color,dcolumn,booktabs,bm}

\graphicspath{{Figures/}} %
\usepackage[colorlinks,
                      citecolor=blue,
                      anchorcolor=red,
                      menucolor=red,
                      linkcolor=red,
                      filecolor=red,
                      runcolor=red,
                      urlcolor=blue,
                      frenchlinks=red]{hyperref}
\begin{document}

\title{Exploration of the hidden charm decays of $Z_{cs}(3985)$}
\author{Qi Wu$^{1,2,3}$}\email{wu$_$qi@pku.edu.cn}
\author{Dian-Yong Chen$^{1,4}$}\email{chendy@seu.edu.cn (corresponding author)}
\affiliation{$^1$School of Physics, Southeast University, Nanjing 210094, China\\
$^2$Center of High Energy Physics, Peking University, Beijing 100871, China\\
$^3$School of Physics and State Key Laboratory of Nuclear Physics and Technology, Peking University, Beijing 100871, China\\
$^4$Lanzhou Center for Theoretical Physics, Lanzhou University, Lanzhou 730000, China}

\begin{abstract}
In the present work, we investigate the  decays of $Z^-_{cs}(3985)$ in the $D_s^{-} D^{\ast0} + D_s^{\ast -}D^0$ molecular frame. By using an effective Lagrangian approach, the branching ratios of $Z^-_{cs}\rightarrow J/\psi K^-$ and $Z^-_{cs}\rightarrow \eta_c K^{\ast-}$ are estimated. It is found that the dominating decay mode of $Z^-_{cs}(3985)$ is open charm channel, while, the branching ratio of the hidden charm decay channels is about $20\%$. In particular, the branching ratios of $Z^-_{cs}\rightarrow J/\psi K^-$ and $Z^-_{cs}\rightarrow \eta_c K^{\ast-}$ are predicted to be an order of $10\%$. Moreover, our estimation indicates that the fraction of the quasi-two-body cascade decay via $Z_{cs}(3985)$ and the total cross section of $e^+ e^- \to K^+ K^- J/\psi$ around 4.68 GeV is $\left(6.8^{+10.5}_{-6.2} \right)\%$. All the predicted branching ratios in the present work could be tested by further measurements in BESIII and Belle II.
\end{abstract}

\date{\today}
\maketitle

\section{Introduction}
\label{sec:introduction}

In the past two decades, the study of exotic hadron states has been a topic of special interest in hadron physics(For recent reviews, we refer to Refs.~\cite{Chen:2016qju,Hosaka:2016pey,Lebed:2016hpi,Esposito:2016noz,Guo:2017jvc,Ali:2017jda,Olsen:2017bmm,Karliner:2017qhf,Yuan:2018inv,Dong:2017gaw,Liu:2019zoy}.). Among the exotic hadron states, the charged charmoniumlike states attracted much attention due to their particular properties and abundant experimental observations. The first charged charmoniumlike states $Z^-_c(4430)$ were observed in the $\pi^-\psi(2S)$ mass spectrum of $B\rightarrow K\pi^-\psi(2S)$ decays by the Belle collaboration in 2007 ~\cite{Belle:2007hrb, Belle:2009lvn, Belle:2013shl}, and then confirmed by the LHCb collaboration with the same process in the year 2014~\cite{LHCb:2014zfx}. Besides $Z_c^-(4430)$, some other charge charmoniumlike states have been observed in the similar $B/\bar{B}$ decay process which benefited to the large bottom meson data sample collected by the Belle and LHCb collaborations, such as $Z_c^+(4050)$ and $Z_c^+(4250)$ in $\bar{B}^0 \to K^- (\pi^+ \chi_{c1}(1P))$~\cite{Belle:2008qeq},  $Z_c^-(4240)$ in $B^0 \to K^+ (\pi^- \psi(2S))$~\cite{LHCb:2014zfx}, $Z_c^+(4200)$ in $\bar{B}^0\to K^- (\pi^+ J/\psi )$~\cite{Belle:2014nuw}, and $Z_c^- (4100)$ in $B^0 \to K^+ (\pi^- \eta_c)$~\cite{LHCb:2018oeg}.

Besides the bottom meson decay, the electron-positron annihilation process provides another unique platform of observing the charged charmoniumlike states; for example, the charmoniumlike state $Z_c^\pm (3900)$ was observed by the BESIII~\cite{Ablikim:2013mio} and Belle~\cite{Liu:2013dau} collaborations in the $\pi^\pm J/\psi$ invariant mass spectrum of $e^+ e^-\rightarrow \pi^+ \pi^- J/\psi$ at $\sqrt{s}=4.26$ GeV in 2013 and then confirmed by CLEO-c collaboration in the same process at $\sqrt{s}=4.17$ GeV~\cite{Xiao:2013iha}, which makes $Z_c(3900)$ the first confirmed charged charmoniumlike state. Later on, another charmoniumlike state $Z^\pm_c(4020)$ was observed by the BESIII collaboration in $e^+ e^-\rightarrow \pi^+ \pi^- h_c$~\cite{Ablikim:2013wzq}. Moreover, these two charged charmoniumlike states have also been observed in the open charm processes ~\cite{BESIII:2013qmu,BESIII:2013mhi} and $\eta_c \rho $ channel~\cite{BESIII:2019rek}.

The rich experimental information of $Z_c(3900)$ and $Z_c(4020)$ makes these two states the most attractive ones among the charge charmoniumlike states. Actually, before the experimental discovering, these kinds of states have been predicted as the charm counterpart of $Z_b(10610)/Z_b(10650)$ or the isospin partner of $X(3872)$ by various methods, such as QCD sum rule~\cite{Chen:2010ze,Chen:2015ata,Chen:2015fsa,Wang:2018ntv,Chen:2013omd,Lee:2008gn,Qiao:2013dda} , potential model~\cite{Liu:2017mrh,Maiani:2007wz,Ebert:2008kb,Patel:2014vua,Hadizadeh:2015cvx,Deng:2015lca,Deng:2014gqa,Zhao:2014qva,Cheung:2007wf,Karliner:2015ina,He:2015mja,
Bugg:2008wu,Swanson:2014tra,Meng:2007fu,Liu:2009wb,Ma:2014zua} and initial single pion emission (ISPE) mechanism ~\cite{Chen:2011xk,Chen:2013wca,Chen:2011pv,Chen:2012yr}. Different from the conventional charmonium, the most possible quark components of charged charmoniumlike states should be $c\bar{c}q\bar{q}$, which indicates that the charged charmoniumlike states could be good candidates of compact tetraquark states. In Refs.~\cite{Maiani:2007wz,Deng:2014gqa,Deng:2015lca,Brodsky:2014xia,Lebed:2017min}, $Z_c(3900)$ and $Z_c(4020)$ were interpreted as a diquark--antidiquark tetraquark state. Moreover, the observed masses of $Z_c(3900)$ and $Z_c(4020)$ were very close to the thresholds of $D^\ast \bar{D}$ and $D^\ast \bar{D}^\ast$, respectively, which motivates the theorists to interpret these two states as deuteronlike molecular states. Using a $D^\ast \bar{D}^\ast$ molecular interpolating current within QCD sum rule, the mass of the molecular state was consistent with $Z_c(4020)$~\cite{Chen:2013omd}. In Refs.~\cite{Dong:2013iqa,Wang:2013cya,Li:2013xia,Li:2014pfa,Xiao:2018kfx,Chen:2016byt,Chen:2015igx}, the decay behavior of $Z_c(3900)$ and $Z_c(4020)$ were investigated in the molecular frame. Besides the QCD exotic interpretations, the structures corresponding to $Z_c(3900)/Z_c(4020)$ have also been reproduced by ISPE mechanism ~\cite{Chen:2013coa,Wang:2013qwa}.

As the SU(3) flavor partner of $Z_c$ states, the existence of hidden charm tetraquark states with strangeness, named $Z_{cs}$, have been investigated after the observation of $Z_c$ states. In Ref.~\cite{Ebert:2008kb}, the hidden charm tetraquark states were evaluated in the relativistic diquark--antidiquark frame and a hidden charm open strange tetraquark state with $J^{P}=1^+$ was predicted with the mass around 4 GeV.  In the molecular scenario, a shallow molecular state composed of $D_s \bar{D}^\ast$ was predicted  in Refs.~\cite{Lee:2008uy,Dias:2013qga}. Similar to the case of $Z_c$ in $e^+ e^- \to \pi^+ \pi^- J/\psi$, some structures with hidden charm and open strange states were also predicted by ISPE mechanism in the $K^+ K^- J/\psi$ decay mode of higher charmonia~\cite{Chen:2013wca}.

In 2014, the Belle collaboration attempted to search $Z_{cs}$ states in the $e^+e^- \to K KJ/\psi$ process, but no obvious structures were observed in the $KJ/\psi$ invariant mass distributions due to the low statistics of the data sample~\cite{Belle:2014fgf}. The experimental breakthrough of observing hidden charm and open strange states appeared when the BESIII collaboration reported a structure, named $Z^-_{cs}(3985)$, in the $K^+$ recoil-mass spectrum of process $e^+ e^-\rightarrow K^+ (D^-_s D^{\ast0}+D^{\ast-}_s D^0)$ at the center-of-mass energy $\sqrt{s}=4.681$ GeV~\cite{Ablikim:2020hsk}. The resonance parameters of $Z_{cs}(3985)$ were reported to be
\begin{eqnarray}
	M &=&(3982.5^{+1.8}_{-2.6}\pm2.1)\ \mathrm{MeV},\nonumber\\
	\Gamma &=&(12.8^{+5.3}_{-4.4}\pm3.0) \ \mathrm{MeV},
\end{eqnarray}
respectively. After the observation of $Z_{cs}(3985)$, the LHCb collaboration made further progress in observing the hidden charm and open strange tetraquark states. Two states, $Z^+_{cs}(4000)$ and $Z^+_{cs}(4220)$,  were reported in the $J/\psi K$ invariant mass spectrum of  $B^+\rightarrow J/\psi\phi K^+$ process~\cite{Aaij:2021ivw}. It should be noticed that the mass of $Z^+_{cs}(4000)$ is consistent with the one of $Z_{cs}(3985)$, but the width of $Z^+_{cs}(4000)$ is much larger than that of $Z_{cs}(3985)$. Thus, it remains to be seen whether $Z_{cs}(4000)$ and $Z_{cs}(3985)$ are the same state.

Similar to $Z_c(3900)/Z_c(4020)$, $Z_{cs}(3985)$ can be interpreted in the tetraquark and molecular scenarios. By virtue of QCD sum rule, the possible configurations of $Z_{cs}(3985)$ were analyzed in the $c\bar{c} q\bar{s}$ tetraquark frame~\cite{Wan:2020oxt, Wang:2020rcx, Wang:2020iqt, Ozdem:2021yvo} and the coupled channels estimation in a constituent quark model~\cite{Jin:2020yjn} and the evaluation in a dynamical diquark model ~\cite{Giron:2021sla} also indicated that the observed $Z_{cs}(3985)$ could be interpreted as a compact tetraquark state. Moreover, the quasidegeneracy between the newly observed $Z_{cs}(3985)$ and $Z_{cs}(4000)$ could be reproduced within tetraquark frame~\cite{Maiani:2021tri}. Considering the mass of $Z_{cs}(3985)$ is very close to the threshold of $D^-_s D^{\ast0}$, it is natural to propose that $Z_{cs}(3985)$ as ${D}^{(\ast)}_s \bar{D}^{(\ast)}$ molecular state~\cite{Meng:2020ihj,Yang:2020nrt,Sun:2020hjw,Wang:2020rcx,Wang:2020htx,Dong:2020hxe,Xu:2020evn,Ozdem:2021yvo,Yan:2021tcp}. However, the calculation from one-boson-exchange (OBE) model does not support a pure hadronic molecular state and there are large compact components in $Z_{cs}$\cite{Liu:2020nge}. Similarly, in Ref.\cite{Chen:2020yvq}, the authors exclude $Z_{cs}$ as a ${D}^{(\ast)}_s \bar{D}^{(\ast)}$ resonance by adopting an OBE model and considering the coupled channel effect. Apart from the above resonance explanations, there also exists some nonresonance explanations, such as a reflection mechanism~\cite{Wang:2020kej}, threshold effect~\cite{Ikeno:2021ptx}.

To date, several charmoniumlike states have been observed near the threshold of $D^\ast \bar{D}_{(s)}$, which are $Z_c(3900)$, $X(3872)$, $Z_{cs}(3985)$ and $Z_{cs}(4000)$. The possible relation between these charmoniumlike states has been discussed in Refs.~\cite{Meng:2020ihj,Yang:2020nrt,Ortega:2021enc} within some approximate symmetries. Considering both $Z_c(3900)$ and $Z_{cs}(3985)$ are narrow states, one can expect that $Z_{cs}(3985)$ is the strange partner of $Z_c(3900)$~\cite{Meng:2020ihj, Yang:2020nrt, Ortega:2021enc}. As for $Z_c(3900)$, it has been observed in the $D^\ast \bar{D}+h.c.$, $J/\psi \pi$ and $\eta_c \rho$ channels. In Ref.\cite{Xiao:2018kfx}, the $J/\psi\pi$ and $\eta_c \rho$ decay modes of $Z_c(3900)$ have been investigated by using triangle loop mechanism and the results supported the molecular interpretation. As pointed out in Ref.~\cite{Xiao:2018kfx}, the estimations of $Z_{c}(3900)$ states via triangle loop mechanism can be used to distinguish the tetraquark and molecular scenario, since the decay mechanism of these two scenarios is different. In the present work, we expand our investigations of the hidden charm decay of $Z_c(3900)$ to its strange partner $Z_{cs}(3985)$ with an effective Lagrangian approach, where $Z_{cs}(3985)$ is assumed to be a $S$-wave $D^{\ast}_s \bar{D}^0/D_s \bar{D}^{\ast}$ molecule state with $I(J^P)=\frac{1}{2}(1^+)$. As indicated in Ref.~\cite{Xiao:2018kfx}, the present estimations will be helpful to distinguish the tetraquark and molecular interpretation of $Z_{cs}(3985)$. Moreover, in the present work, the decay mode $\eta_c K^\ast$ will be predicted, which could be tested by further experimental measurements in BESIII and Belle II.

This paper is organized as follows. After the Introduction, we present the model used in the estimations of the hidden charm decays of $Z_{cs}(3985)$. The numerical results and discussions are presented in  Sec.~\ref{Sec:Num}, and Sec.~\ref{sec:summary} is devoted to a short summary.

\section{Hidden charm decay of $Z_{cs}(3985)$ }
\label{sec:model}

\begin{figure}[htb]
\begin{tabular}{ccc}
  \centering
 \includegraphics[width=4.2cm]{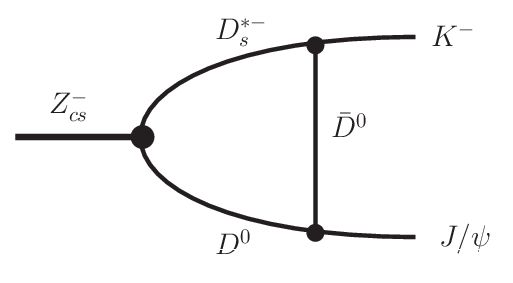}&
 \includegraphics[width=4.2cm]{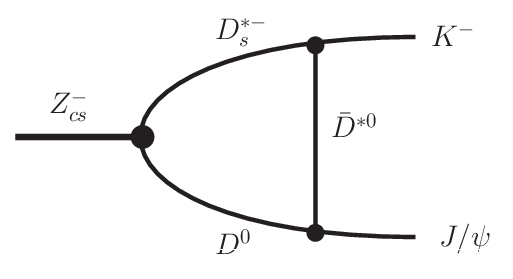}\\
 \\
 $(a)$ & $(b)$ \\
 \\
 \includegraphics[width=4.2cm]{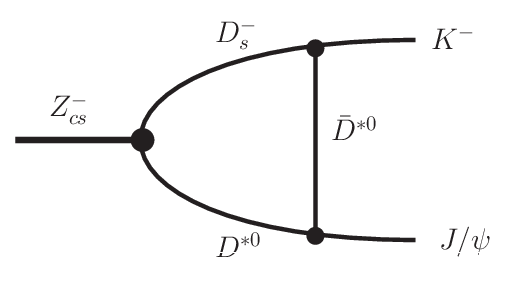}&
 \includegraphics[width=4.2cm]{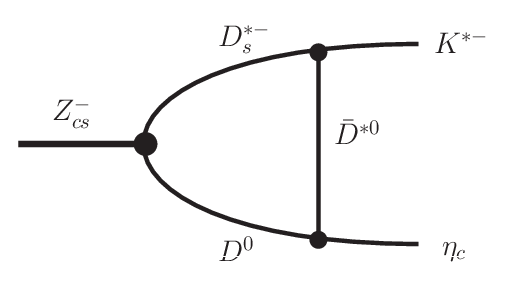}&\\
 \\
 $(c)$ & $(d)$&\\
  \\
 \includegraphics[width=4.2cm]{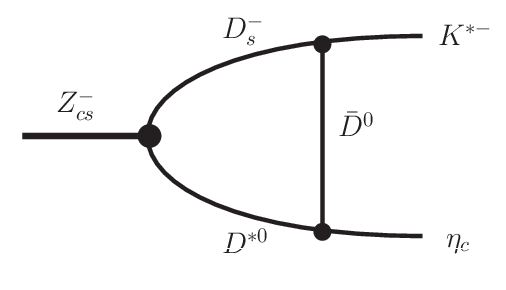}&
 \includegraphics[width=4.2cm]{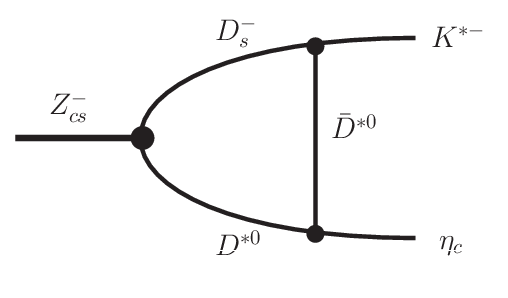}&\\
 \\
 $(e)$ & $(f)$&\\
 \end{tabular}
  \caption{The typical diagrams  contributing to $Z^-_{cs} \rightarrow J/\psi K^-$(a--c) and $Z^-_{cs} \rightarrow \eta_c K^{\ast-}$(d--f); other similar diagrams could be obtained by replacing $D^{(\ast)}_s$ and $D^{(\ast)}$ with $D^{(\ast)}$ and $D^{(\ast)}_s$, respectively.}\label{Fig:Tri}
\end{figure}

In the present work, we assume that $Z^-_{cs}$ is a $S$-wave molecule with $I(J^P)=\frac{1}{2}(1^+)$ given by the superposition of $D^{\ast-}_s D^0$ and $D^-_s D^{\ast0}$ hadronic configurations, which is,
\begin{eqnarray}
|Z^-_{cs}\rangle &=& \frac {1} {\sqrt {2}} (|D^{\ast-}_s D^0\rangle+|D^-_s D^{\ast0}\rangle).
\label{eq:wavef}
\end{eqnarray}
As a result, one can depict the effective coupling of $Z^-_{cs}$ to its components in terms of the following effective Lagrangian,
\begin{eqnarray}
{\cal L}_{Z_{cs}}&=&\frac {g_{Z_{cs}}} {\sqrt {2}}Z^{\dagger\mu}_{cs}\left( \bar{D}^{\ast}_{s\mu} D + \bar{D}_s D^{\ast}_\mu \right ),
\label{eq:lagX}
\end{eqnarray}
where $g_{Z_{cs}}$ is the effective coupling constant.

As for the interactions between the charmonia and charmed mesons, the effective Lagrangians could be constructed based on heavy quark limit, and the relevant effective Lagrangians read~\cite{Oh:2000qr,Colangelo:2002mj,Casalbuoni:1996pg}
\begin{eqnarray}
\mathcal{L}_{\psi D^{(\ast)}D^{(\ast)}}&=&-ig_{\psi DD}\psi_\mu(\partial^\mu DD^\dag-D\partial^\mu D^\dag)\nonumber\\
&&+g_{\psi D^{\ast}D}\varepsilon^{\mu\nu\alpha\beta}\partial_\mu \psi_\nu(D^\ast_\alpha \stackrel{\leftrightarrow}{\partial}_\beta D^\dag-D \stackrel{\leftrightarrow}{\partial}_\beta D_\alpha^{\ast\dag})\nonumber\\
&&+ig_{\psi D^{\ast}D^{\ast}}\psi^\mu(D^\ast_\nu \stackrel{\leftrightarrow}{\partial}^\nu D^{\ast\dag}_\mu\nonumber\\
&&+D^\ast_\mu \stackrel{\leftrightarrow}{\partial}^\nu D^{\ast\dag}_\nu-D^\ast_\nu \stackrel{\leftrightarrow}{\partial}_\mu D^{\ast\dag\nu}),\nonumber\\
\mathcal{L}_{\eta_c D^{\ast}D^{(\ast)}}&=&-ig_{\eta_c D^{\ast}D}\eta_c(D \stackrel{\leftrightarrow}{\partial}_\mu D^{\ast\dag\mu}+D^{\ast\mu}\stackrel{\leftrightarrow}{\partial}_\mu D^\dag)\nonumber\\
&&-g_{\eta_c D^{\ast}D^{\ast}}\varepsilon^{\mu\nu\alpha\beta}\partial_\mu \eta_c D^{\ast}_\nu \stackrel{\leftrightarrow}{\partial}_\alpha D^{\ast\dag}_\beta.
\label{eq:charmonium}
\end{eqnarray}

The Lagrangians relevant to the light vector and pseudoscalar mesons can be constructed based on the heavy quark limit and chiral
symmetry~\cite{Casalbuoni:1996pg,Colangelo:2003sa,Cheng:2004ru}, which are
\begin{eqnarray}
 {\cal L} &=& -ig_{D^{\ast }D
{\mathcal P}}\left( D^{\dag}_i \partial_\mu {\mathcal P}_{ij} D_j^{\ast \mu}-D_i^{\ast \mu\dagger} \partial_\mu {\mathcal P}_{ij}  D_j\right) \nonumber \\
&& +\frac{1}{2}g_{D^\ast D^\ast {\mathcal P}}\varepsilon _{\mu
\nu \alpha \beta }D_i^{\ast \mu \dag}\partial^\nu {\mathcal P}_{ij}  {\overset{
\leftrightarrow }{\partial }}{\!^{\alpha }} D_j^{\ast \beta } - ig_{{D}{D}\mathcal{V}} {D}_i^\dagger {\stackrel{\leftrightarrow}{\partial}}{\!_\mu} {D}^j(\mathcal{V}^\mu)^i_j \nonumber \\
&& -2f_{{D}^*{D}\mathcal{V}} \epsilon_{\mu\nu\alpha\beta}
(\partial^\mu \mathcal{V}^\nu)^i_j
({D}_i^\dagger{\stackrel{\leftrightarrow}{\partial}}{\!^\alpha} {D}^{*\beta j}-{D}_i^{*\beta\dagger}{\stackrel{\leftrightarrow}{\partial}}{\!^\alpha} {D}^j) \nonumber
\\
&&+ ig_{{D}^*{D}^*\mathcal{V}} {D}^{*\nu\dagger}_i {\stackrel{\leftrightarrow}{\partial}}{\!_\mu} {D}^{*j}_\nu(\mathcal{V}^\mu)^i_j \nonumber \\
&& +4if_{{D}^*{D}^*\mathcal{V}} {D}^{*\dagger}_{i\mu}(\partial^\mu \mathcal{V}^\nu-\partial^\nu
\mathcal{V}^\mu)^i_j {D}^{*j}_\nu +{\rm H.c.} , \label{eq:light-meson}
 \label{eq:LDDV}
 \end{eqnarray}
where the ${D}^{(\ast)\dagger}=(\bar{D}^{(\ast)0},D^{(\ast)-},D^{(\ast)-}_s)$ are the charmed
meson triplets, $\mathcal P$ and ${\mathcal V}_\mu$ are $3\times 3$ matrices form of pseudoscalar and vector mesons, and their concrete forms are
\begin{eqnarray}
\mathcal{P} &=&
\left(\begin{array}{ccc}
\frac{\pi^{0}}{\sqrt 2}+\alpha\eta+\beta\eta\prime &\pi^{+} &K^{+}\\
\pi^{-} &-\frac{\pi^{0}}{\sqrt2}+\alpha\eta+\beta\eta \prime&K^{0}\\
K^{-} &\bar K^{0} &\gamma\eta+\delta\eta\prime
\end{array}\right),\nonumber\\ \nonumber\\
\mathcal{V} &=& \left(\begin{array}{ccc}\frac{\rho^0} {\sqrt {2}}+\frac {\omega} {\sqrt {2}}&\rho^+ & K^{*+} \\
\rho^- & -\frac {\rho^0} {\sqrt {2}} + \frac {\omega} {\sqrt {2}} & K^{*0} \\
K^{*-}& {\bar K}^{*0} & \phi \\
\end{array}\right) ,
\end{eqnarray}
with $\alpha$ and $\beta$ being parameters related to the mixing angle $\theta$. All the relevant coupling constants will be discussed in the following section.

With the above effective Lagrangians, we can obtain the amplitudes for $Z_{cs}\rightarrow J/\psi K$ corresponding to diagrams Fig.~\ref{Fig:Tri}(a)--~\ref{Fig:Tri}(c), which are
\begin{eqnarray}
\mathcal{M}_{a}&=&i^3 \int\frac{d^4 q}{(2\pi)^4}\Big[g_{Z_{cs}} \epsilon_{0\mu}\Big]\Big[-ig_{D^\ast DP}(-)ip_{3\nu}\Big]\nonumber\\
&&\times\Big[-ig_{\psi DD}\epsilon_{4 \alpha}t (-ip^\alpha_2+iq^\alpha)\Big]\nonumber\\
&&\frac{-g^{\mu\nu}+p^\mu_1 p^\nu_1 /m^2_1}{p^2_1-m^2_1}\frac{1}{p^2_2-m^2_2}\frac{1}{q^2-m^2_q}\mathcal{F}^3(q^2,m_q^2)\nonumber 
\end{eqnarray}
\begin{eqnarray}
\mathcal{M}_{b}&=&i^3 \int\frac{d^4 q}{(2\pi)^4}\Big[g_{Z_{cs}} \epsilon_{0\kappa}\Big]\Big[\frac{1}{2}g_{D^\ast D^\ast P}\varepsilon_{\mu\nu\alpha\beta}ip^\nu_3 (-ip^\alpha_1-iq^\alpha)\Big]\nonumber\\
&&\Big[g_{\psi D^\ast D}\varepsilon_{\rho\sigma\tau\xi}ip^\rho_4 \epsilon^\sigma_{4}(-)(iq^\xi-ip^\xi_2)\Big]\frac{-g^{\kappa\beta}+p^\kappa_1 p^\beta_1 /m^2_1}{p^2_1-m^2_1}\nonumber\\
&&\frac{1}{p^2_2-m^2_2}\frac{-g^{\mu\tau}+q^{\mu}q^{\tau}/m^2_q}{q^2-m^2_q}\mathcal{F}^3(q^2,m_q^2)\nonumber\\
\mathcal{M}_{c}&=&i^3 \int\frac{d^4 q}{(2\pi)^4}\Big[g_{Z_{cs}} \epsilon_{0\rho}\Big]\Big[-ig_{D^\ast DP}ip_{3\sigma}\Big]\Big[ig_{\psi D^\ast D^\ast}\epsilon^\tau_{4}(-i)\nonumber\\
&&\Big(g_{\xi\theta}g_{\lambda\tau}(p^\theta_2-q^\theta)+g_{\xi\tau}g_{\lambda\theta}(p^\theta_2-q^\theta)-g_{\xi\theta}g^\theta_{\lambda}(p_{2\tau}
-q_{\tau})\Big)\Big]\nonumber\\
&&\frac{1}{p^2_1-m^2_1}\frac{-g^{\rho\lambda}+p^{\rho}_2 p^{\lambda}_2 /m^2_2}{p^2_2-m^2_2}\frac{-g^{\sigma\xi}+q^{\sigma} q^{\xi} /m^2_q}{q^2-m^2_q}\mathcal{F}^3(q^2,m_q^2),\nonumber\\ \label{Eq:amp1}
\end{eqnarray}

The amplitudes for $Z_{cs}\rightarrow \eta_c K^\ast$ corresponding to diagrams Fig.~\ref{Fig:Tri}(d)--~\ref{Fig:Tri}(f) are
\begin{eqnarray}
\mathcal{M}_{d}&=&i^3 \int\frac{d^4 q}{(2\pi)^4}\Big[g_{Z_{cs}} \epsilon_{0\mu}\Big]\Big[ig_{D^\ast D^\ast V}g^\nu_\tau g_{\theta\nu}(iq_\kappa+ip_{2\kappa})\epsilon^\kappa_{3}\nonumber\\
&&+4if_{D^\ast D^\ast V}g_{\tau\kappa}g_{\theta\nu}i(p^\kappa_3 \epsilon^\nu_{3}-p^\nu_3 \epsilon^\kappa_{3})\Big]\nonumber\\
&&\Big[-ig_{\eta_c D^\ast D}(-iq_{\sigma}+ip_{2\sigma})\Big]\frac{-g^{\mu\tau}+p^{\mu}_1 p^{\tau}_1 /m^2_1}{p^2_1-m^2_1}\nonumber\\
&&\frac{1}{p^2_2-m^2_2}\frac{-g^{\theta\sigma}+q^{\theta} q^{\sigma} /m^2_q}{q^2-m^2_q}\mathcal{F}^3(q^2,m_q^2),\nonumber\\
\mathcal{M}_{e}&=&i^3 \int\frac{d^4 q}{(2\pi)^4}\Big[g_{Z_{cs}} \epsilon_{0\mu}\Big]\Big[-ig_{DDV}(-ip_{1\nu}-iq_{\nu})\epsilon^\nu_{3}\Big]\nonumber\\
&&\Big[-ig_{\eta_c D^\ast D}(-ip_{2\sigma}+iq_{\sigma})\Big]\frac{1}{p^2_1-m^2_1}\nonumber\\
&&\frac{-g^{\mu\sigma}+p^{\mu}_2 p^{\sigma}_2 /m^2_2}{p^2_2-m^2_2}\frac{1}{q^2-m^2_q}\mathcal{F}^3(q^2,m_q^2),\nonumber\\
\mathcal{M}_{f}&=&i^3 \int\frac{d^4 q}{(2\pi)^4}\Big[g_{Z_{cs}} \epsilon_{0\kappa}\Big]\Big[-2f_{D^\ast DV}\varepsilon_{\mu\nu\alpha\beta}ip^\mu_3 \epsilon^\nu_{3}(-)\nonumber\\
&&(-ip^\alpha_1-iq^\alpha)\Big]\Big[-g_{\eta_c D^\ast D^\ast}\epsilon_{\rho\sigma\tau\xi}ip^\rho_4 (-ip_{2\tau}+iq_{\tau})\Big]\nonumber\\
&&\frac{1}{p^2_1-m^2_1}\frac{-g^{\kappa\xi}+p^{\kappa}_2 p^{\xi}_2 /m^2_2}{p^2_2-m^2_2}\frac{-g^{\beta\sigma}+q^{\beta} q^{\sigma} /m^2_q}{q^2-m^2_q}\mathcal{F}^3(q^2,m_q^2),\nonumber\\ \label{Eq:amp2}
\end{eqnarray}
where $p_1(m_1)$, $p_2(m_2)$, and $q(m_q)$ are the momentum (mass) of the intermediate mesons, while $p_0(m_0,\ \epsilon_0)$, $p_3(m_3,\ \epsilon_3)$, and $p_4(m_4,\ \epsilon_4)$ are the momentum (mass, polarization vector) of the initial $Z_{cs}(3985)$, the light meson, and the charmonia in final states.

In the above amplitudes, a form factor is introduced to represent the off-shell effect of the exchanging charmed mesons, and the form factor also plays the role of avoiding integration divergent. In the present work, we adopt the form factor in a monopole form, which is
\begin{eqnarray}
\mathcal{F}\left(q^{2}\right)=\frac{m^{2}-\Lambda^{2}}{q^{2}-\Lambda^{2}},\label{Eq:FFs1}
\end{eqnarray}
where the parameter $\Lambda$ can be further reparametrized as $\Lambda_{D^{(\ast)}}=m_{D^{(\ast)}}+\alpha\Lambda_{\rm QCD}$ with $\Lambda_{\rm QCD}=0.22 \ {\rm GeV}$, and $m_{D^{(\ast)}}$ is the mass of the exchanged charmed meson. The model parameter $\alpha$ should be of order of unity~\cite{Tornqvist:1993vu,Tornqvist:1993ng,Locher:1993cc,Li:1996yn}, but its concrete value cannot be estimated by the first principle. In practice, the value of $\alpha$ is usually determined by comparing theoretical estimates with the corresponding experimental measurements.

\section{Numerical Results and discussions}
\label{Sec:Num}
\subsection{Coupling constants}

In the heavy quark limit, the coupling constants of $S$-wave charmonia and charmed or charmed-strange meson pairs can relate to the gauge coupling $g_1$ by~\cite{Oh:2000qr,Colangelo:2002mj,Casalbuoni:1996pg}
\begin{eqnarray}
g_{\psi DD} &=& 2g_1 \sqrt{m_{\psi}}m_D,\nonumber\\  g_{\psi D^\ast D} &=& 2g_1 \sqrt{m_D m_{D^\ast}/m_{\psi}},\nonumber \\
g_{\psi D^\ast D^\ast} &=& 2g_1 \sqrt{m_{\psi}}m_{D^\ast},\nonumber \\
g_{\eta_c D^*D}&=&2 g_1 \sqrt{m_{\eta_c} m_D m_{D^*}}, \nonumber \\
g_{\eta_c D^* D^*} &=& 2g_1 m_{D^*}/\sqrt{m_{\eta_c}},
\end{eqnarray}
where $g_1=\sqrt{m_\psi}/(2m_D f_\psi)$ and $f_\psi=426 {\rm MeV}$ are the $J/\psi$ decay constants.

Considering heavy quark limit and chiral symmetry, the  coupling constants relevant to the light vector and pseudoscalar mesons are ~\cite{Casalbuoni:1996pg,Cheng:2004ru}
\begin{eqnarray}
g_{{ D}{ D}V} = g_{{ D}^*{ D}^*V}=\frac{\beta g_V}{\sqrt{2}} , \quad f_{{ D}^*{ D}V}=\frac{ f_{{ D}^*{ D}^*V}}{m_{{ D}^*}}=\frac{\lambda g_V}{\sqrt{2}} \nonumber\, , \\
g_{{D}^{*} {D} \mathcal{P}}=\frac{2 g}{f_{\pi}} \sqrt{m_{{D}} m_{{D}^{*}}}, \quad g_{{D}^{*} {D}^{*} {P}}=\frac{g_{{ D}^{*} { D} {\mathcal {P}}}}{\sqrt{m_{{D}} m_{{D}^{*}}}},
\end{eqnarray}
where the parameter $\beta=0.9$, $g_V = {m_\rho /f_\pi}$ with $f_\pi = 132$ MeV being the decay constant of pion~\cite{Casalbuoni:1996pg}. By matching the form factor obtained from the light cone sum rule and that calculated from lattice QCD, one can obtain the parameter $\lambda = 0.56 \, {\rm GeV}^{-1} $ and $g=0.59$~\cite{Isola:2003fh}.

Besides the above coupling constants, the one of $Z_{cs}(3985)$ and its components as given in Eq.~(\ref{eq:lagX}) is undetermined. As for $g_{Z_c D^\ast D}$, it was determined by the relevant partial width of $Z_c(3900) \to D^\ast \bar{D}$. However, there is no direct experimental measurement of the partial width of $Z_{cs}(3985) \to D_s^\ast \bar{D}$ so far. Here, we assume that $D_s^\ast \bar{D}+h.c.$, $J/\psi K$, and $\eta_c K^\ast$ are the dominant decay modes of $Z_{cs}(3985)$ and then the coupling constant $g_{Z_{cs} D_s^\ast D}$ can be determined by the total widths of  $Z_{cs}(3985)$. In the same time, we can obtain the branching ratios of these three channels with such assumptions.

\subsection{Branching ratios}

\begin{table*}[htb]
\centering
\caption{The branching ratios of $Z^-_{cs}\rightarrow J/\psi K^-$, $Z^-_{cs}\rightarrow \eta_c K^{\ast-}$ and $Z^-_{cs}\rightarrow D^0 D^{\ast-}_s+c.c.$(in unites of $\%$) depending on the model parameter $\alpha$.
 \label{Table:br}}
 \renewcommand\arraystretch{2}
\begin{tabular}{p{3cm}<\centering p{2cm}<\centering p{2cm}<\centering p{2cm}<\centering p{2cm}<\centering p{2cm}<\centering}
  \toprule[1pt]
  $\alpha$         & 3 & 3.5 & 4 & 4.5 & 5 \\
  \midrule[1pt]
  $\mathcal{B}[Z^-_{cs}\rightarrow J/\psi K^-]$       & 1.3 & 2.4 & 4.0 & 6.0 & 8.3  \\
  $\mathcal{B}[Z^-_{cs}\rightarrow \eta_c K^{\ast-}]$  & 0.9 & 2.1 & 4.2 & 7.6 & 12.5  \\
  $\mathcal{B}[Z^-_{cs}\rightarrow D^0 D^{\ast-}_s+c.c.]$  & 97.8 & 95.5 & 91.8 & 86.4 & 79.2 \\
  \bottomrule[1pt]
\end{tabular}
\end{table*}

With the decay amplitudes in the above section, one can estimate the partial width of the hidden charm decay channels, which are proportional to $g_{Z_{cs}}^2$ and depending on the model parameter $\alpha$, i.e,
\begin{eqnarray}
\Gamma_{Z_{cs} \to J/\psi K} =g_{Z_{cs}}^2 f_{J/\psi K}(\alpha),\nonumber\\
\Gamma_{Z_{cs} \to \eta_c K^\ast} =g_{Z_{cs}}^2 f_{\eta_c K^\ast }(\alpha),
\end{eqnarray}
where $f_{J/\psi K}(\alpha)$ and $f_{\eta_c K^\ast }(\alpha)$ are functions of $\alpha$, which can be estimated from the decay amplitudes listed in Eqs. (\ref{Eq:amp1}), (\ref{Eq:amp2}). In Ref.~\cite{Xiao:2018kfx}, the hidden-charm decays of $Z^{(\prime)}_c$ were estimated with $\alpha=3\sim5.5$, and we found that the measured partial width of $Z_c\rightarrow J/\psi \pi$ could be reproduced in the range of $\alpha=3.63\sim4.75$. Considering the similarity between $Z_c(3900)$ and $Z_{cs}(3985)$, we vary the model parameter $\alpha$ from $3$ to $5$.

As for $D_s^\ast \bar{D} +h.c.$ channel, the partial widths can be directly estimated from the effective Lagrangian listed in Eq. (\ref{eq:lagX}), which is
\begin{eqnarray}
\Gamma_{Z_{cs} \to D_s^\ast \bar{D}+h.c}&=& \frac{1}{3}\frac{1}{8\pi}\frac{|\vec{p}|}{m^2}\frac{g_{Z_{cs}}^2}{2}\Big(-g^{\mu\nu}+\frac{p^\mu p^\nu}{m^2}\Big)\nonumber\\
&&\times\Big(\Big(-g^{\mu\nu}+\frac{p^\mu_1 p^\nu_1}{m^2_1}\Big)+\Big(-g^{\mu\nu}+\frac{p^\mu_2 p^\nu_2}{m^2_2}\Big)\Big)\nonumber
\\
\end{eqnarray}

As discussed above, the partial widths of $Z_{cs}(3985) \to D_s^\ast \bar{D} +h.c.,\ J/\psi K$, and $\eta_c K^\ast$ are all proportional to $g_{Z_{cs}}^2$. With the assumption that these three channels are the dominant decay channels of $Z_{cs}$, i.e., $\Gamma_{Z_{cs} \to D_s^\ast \bar{D}+h.c.} +\Gamma_{Z_{cs} \to J/\psi K} +\Gamma_{Z_{cs} \to \eta_c K^\ast} \simeq \Gamma_{Z_{cs}}^\mathrm{Total}$, one can estimate the coupling constant $g_{Z_{cs}}$ depending on the model parameter $\alpha$. In particular, the coupling constant $g_{Z_{cs}}$ decreases from $6.7$ to $6.0$ when $\alpha$ increases from $3$ to $5$ by using the center value of the measured width.

With the determined $g_{Z_{cs}}$, one can further obtain the branching ratios of $Z^-_{cs}\rightarrow J/\psi K^-$, $Z^-_{cs}\rightarrow \eta_c K^{\ast-}$ and $Z_{cs}^- \to D_s^{\ast-} D^0 +D_s^- D^0$, which are listed in Table~\ref{Table:br}. Similar to $Z_c(3900)$, our estimations indicate $Z_{cs}^-(3985)$ dominantly decays to its component mesons, i.e $D_s^{\ast-} D^0 +D_s^- D^0$. As for the hidden charm channels, the branching ratios can reach up to $20\%$. As for $Z_c(3900)$, the ratio of the fractions of $D^\ast \bar{D} $ and $J/\psi \pi$ channels were measured to be $6.2 \pm 1.1 \pm 2.7$, while the ratio of fractions of $\eta_c \rho$ and $J/\psi \pi$ channels were $2.1 \pm 0.8 $. From these experimental measurements, one can find that the branching ratios for $J/\psi \pi$ and $\eta_c \rho$ are about $6.5\%$ and $13.5\%$, respectively. As shown in Table~\ref{Table:br}, the estimated branching ratios of hidden charm decay modes are comparable to those of $Z_c(3900)$, which reflect the similarity between $Z_c(3900)$ and $Z_{cs}(3985)$. As listed in Table~\ref{Table:br}, the branching ratios of the hidden charm decay modes increase with the parameter $\alpha$, and the $\alpha$ dependences of these two branching ratios are similar; thus, one can expect that their ratio should weakly depend on the parameter $\alpha$. In the present work, the ratio of $\mathcal{B}(Z_{cs} \to \eta_c K^\ast)$ and $\mathcal{B}(Z_{cs} \to J/\psi K)$ is estimated to be $0.69 \sim 1.52$, which is also comparable with the measured one for $Z_c(3900)$.

From the above estimations and discussions, one can find that the decay behavior of $Z_{cs}(3985)$ is very similar to the one of $Z_c(3900)$ when we assume $Z_{cs}(3985)$ as a $D_s^{\ast-} D^0 +D_s^{-}D^{\ast 0}$ molecular state. The branching ratios of the hidden charm decay channels are similar to those of $Z_c(3900)$. Thus, with the accumulation of the experimental data, the charmoniumlike states $Z_{cs}(3985)$ are expected to be observed in the hidden charm decay channels by further measurements of BESIII and Belle II. It should be noticed that the BESIII collaboration measured the product of the Born cross sections $\sigma [e^+ e^-\rightarrow K^+  Z^-_{cs}+c.c.]$ and the branching fractions for $Z^-_{cs}\to  D^-_s D^{\ast 0}+D_s^{\ast-} D^0$ at $\sqrt{s}=4.681$ GeV to be~\cite{Ablikim:2020hsk}
\begin{eqnarray}
&&\sigma [e^+ e^-\rightarrow K^+  Z^-_{cs}+c.c]\times \mathcal{B}[Z^-_{cs}\to  D^-_s D^{\ast0}+D_s^{\ast-} D^0] \nonumber\\&=& \left(4.4^{+0.9}_{-0.8}\pm 1.4\right)\ \mathrm{pb}.
\end{eqnarray}

In the present work, the ratio of $\mathcal{B}[Z^-_{cs}\rightarrow J/\psi K^-]$ and $\mathcal{B}[Z^-_{cs}\rightarrow D^0 D^{\ast-}_s+c.c.]$ is estimated to be
\begin{eqnarray}
\frac{\mathcal{B}[Z^-_{cs}\rightarrow J/\psi K^-]}{\mathcal{B}[Z^-_{cs}\to  D^-_s D^{\ast0}+D_s^{\ast-} D^0] }=\left(4.4^{+6.1}_{-3.0} \right) \%,
\end{eqnarray}
where the center value is determined with $\alpha=4$, while the upper and lower errors are estimated with $\alpha=5$ and $\alpha=3$, respectively. Considering that the width of $Z_{cs}$ is small, we can have the following approximate relation:
\begin{eqnarray}
&&\sigma[e^+ e^-\rightarrow K^+  Z^-_{cs}  +c.c\rightarrow K^+ K^- J/\psi]\nonumber\\
&\simeq& \sigma [e^+ e^-\rightarrow K^+ Z^-_{cs} +c.c] \times \mathcal{B}[Z^-_{cs}\rightarrow J/\psi K^-] \nonumber\\
&=& \left(0.19^{+0.28}_{-0.15} \right) \ {\rm pb}.
\end{eqnarray}
In Ref.~\cite{Belle:2014fgf}, the Belle collaboration measured the cross sections for $e^+ e^-\rightarrow K^+ K^- J/\psi$ at the center-of-mass energies 4.675 GeV, which is $2.8^{+1.4}_{-1.2}\pm 0.3$ pb. With this cross section, we can roughly estimate the fraction of the quasi-two-body cascade decay via $Z_{cs}(3985)$ and the total cross section around 4.68 GeV to be
\begin{eqnarray}
\frac{\sigma[e^+ e^-\rightarrow K^+ Z^-_{cs}+c.c. \rightarrow K^+ K^- J/\psi]}{\sigma [e^+ e^-\rightarrow K^+ K^- J/\psi]}=\left(6.8^{+10.5}_{-6.2} \right) \%,
\end{eqnarray}
which could be tested by further measurements in BESIII and Belle II.

\section{Summary}
\label{sec:summary}
Benefit from the improvement of experimental techniques and the accumulation of experimental data, increasing charmoniumlike states have been observed experimentally. Among these charmoniumlike states, the ones with charge attracted particular attention because of their exotic nature. Recently, BESIII collaboration observed a charged charmoniumlike state $Z^-_{cs}(3985)$ in the $K^+$ recoil-mass spectrum of process $e^+ e^-\rightarrow K^+ (D^-_s D^{\ast0}+D^{\ast-}_s D^0)$, which could be assigned as a strange partner of $Z_c(3900)$.

Inspired by the observations from the BESIII collaboration, we study the hidden charm decays of $Z^-_{cs}(3985)$ via triangle loop mechanism by using an effective Lagrangian approach. The branching ratios of $Z^-_{cs}\rightarrow J/\psi K^-$ and $Z^-_{cs}\rightarrow \eta_c K^{\ast-}$ are predicted to be of order $10\%$, and the dominant decay mode of $Z^-_{cs}(3985)$ is an open charm channel. Our estimations indicate the decay property of $Z_{cs}(3985)$ is in line with the one of $Z_c(3900)$, which supports $Z^-_{cs}(3985)$ as a molecular state.

Moreover, based on the available experimental data and the present estimation, we find the fraction of the quasi-two-body cascade decay via $Z_{cs}(3985)$ and the total cross section of $e^+ e^- \to K^+ K^- J/\psi$ around 4.68 GeV is $\left(6.8^{+10.5}_{-6.2} \right)\%$, which could be tested by further measurements in BESIII and Belle II.

\section*{Acknowledgement}
This work is supported by the National Natural Science Foundation of China (NSFC) under Grant No.11775050.

\end{document}